\documentstyle[aps,12pt]{revtex}

\begin{document}
\title{ Perturbative Analysis of Bianchi IX using Ashtekar Formalism}
\author{Esteban A. Calzetta\thanks{%
Electronic address: calzetta@iafe.uba.ar}$^{1,2}$, Marc Thibeault\thanks{%
Electronic address: marc@iafe.uba.ar}$^1$}
\address{$^1$Instituto de Astronom\'\i a y F\'\i sica del Espacio, \\
c.c. 67, suc 28 (1428) Buenos Aires Argentina and \\
$^2$Departamento de F\'\i sica, Facultad de Ciencias Exactas y Naturales, \\
Ciudad Universitaria (1428) Buenos Aires, Argentina}
\maketitle

\begin{abstract}
The goal of this paper is to provide a new analysis of the classical
dynamics of Bianchi type I, II and IX models by applying conventional
Hamiltonian methods in the language of Ashtekhar variables. We show that
Bianchi type II models can be seen as a perturbation of Bianchi I ones, and
integrated. Bianchi IX models can be seen, in turn, as a perturbation of
Bianchi IIs, but here the integration algorithm breaks down. This is an
''interesting failure'', bringing light onto the chaotic nature of Bianchi
type IX dynamics.As a by product of our analysis we filled some gaps in the
literature, such us recovering the BKL map in this context.
\end{abstract}

\section{Introduction}

The goal of this paper is to provide a new analysis of the classical
dynamics of Bianchi type I, II and IX models by applying conventional
Hamiltonian methods in the language of Ashtekhar variables. We show that
Bianchi type II models can be seen as a perturbation of Bianchi I ones, and
integrated. Bianchi IX models can be seen, in turn, as a perturbation of
Bianchi IIs, but here the integration algorithm breaks down. This is an
''interesting failure'', bringing light onto the chaotic nature of Bianchi
type IX dynamics.

Bianchi models can be traced back to the nineteen century\cite{Bianchi}, at
least in the abstract, but their use in physical considerations had of
course to wait for the invention of General Relativity (GR) and the
development of cosmology. Bianchi I was first discussed by Kasner \cite
{Kasner} and further by Taub\cite{Taub}, Misner \cite{Misner} and Lifshitz
and Khalatnikov\cite{Lifshitz}.The empty Bianchi II model was solved first
by Taub in a work where he discussed empty spaces with a three parameter
group of motion\cite{Taub}. Misner reformulated homogeneous cosmology using
Hamiltonian methods\cite{Misner2}, after the general hamiltonian formulation
due to Arnowitt, Deser and Misner (ADM)\cite{ADM}. This led Misner to
introduce the fruitful concept of minisuperspace \cite{Misner3} where the
cosmological evolution was equivalent to the motion of a particle in a
potential. The cosmological implications of Bianchi IX were treated by Misner%
\cite{Misner4} and by Belinski, Lifshitz and Khalatnikov (BKL)\cite{BKL}. We
should note that when we write Bianchi IX, we restrict ourselves to the
empty, diagonal case.\cite{Ryan-Shepley}

Bianchi IX is a homogeneous solution that became popular as a generic model
of the approach to the singularity, even in inhomogeneous models \cite
{Lifshitz}. Although this is a controversial point \cite{Barrow}, the
approach of treating the inhomogeneous case near the singularity as a
perturbation (where the inhomogeneity were taking un account by a
development in spatial derivatives ) from a zero order homogeneous solution
(Bianchi IX) has been fruitfull \cite{Deruelle}. In any case, the approach
to the singularity as viewed by BKL shed new light on the dynamics of this
complex solution since it can for be appproximated as a one dimensional map.
During most of its evolution, Bianchi IX can be seen as a sequence of Kasner
solutions, with the parameters of one Kasner "era" being mapped into those
of the next; Barrow proved this map to be chaotic\cite{Barrow2}.

Actually, this was not the first time that chaos was proven in an
approximation of Bianchi IX. Chitre \cite{Chitre} proved that the Misner
approximation of Bianchi IX was chaotic analitically using a know theorems
about geodesic flows and the hyperbolic plane (see the article of Misner\cite
{Misner5} for a summary). Zardecki \cite{Zardecki} first made numerical
integrations of the Einstein Equations describing the Binchi IX evolution in
the BKL form and obtained results that seemed to agree with these analysis.
However, further investigations by Francisco and Matsas \cite{Francisco} and
Rugh \cite{Rugh} showed that the initial conditions posed by Zardecki did
not obey the scalar constraint. This could introduce negative energy
densities, preventing collapse at the singularity and setting up chaotic
motions. Furthermore, Francisco and Matsas computed the first Lyapunov
exponent and showed that it tended toward zero. Analytical results from Burd
et al.\cite{Burd} and Hobill\cite{Hobill} showed that the Lyapunov exponents
had to be identically zero. Pullin and Rugh\cite{Pullin} \cite{Rugh2}
explained this result by proving that the use of different time
parametrizations could produce different Lyapunov exponents. The situation
became confused, as usual chaos indicators seemed unable to capture the
''irregular '' character of (deterministic) Bianchi IX. By example a first
analysis using The Painlev\'e test seemed to indicate that it was integrable%
\cite{CGR} but this claim was contested shortly thereafter\cite{LMC}\cite
{CGR2} using the perturbative Painlev\'e test. Again, these works strongly
indicated that the Mixmaster universe is indeed chaotic but we ought to
remember that neither the Painlev\'e test nor the perturbative one have
reached theorem status, that is, at this time, they can only be taken as
chaotic behavior indicators. The latest original indication toward the
chaotic character of the Mixmaster was given by Cornish and Levin \cite{C-L}%
. They uncovered a (multi-)fractal repellor both in a two parameter map, the
so-called Farey map\cite{Chernoff}, and in the full continuous dynamics.
Fractality could be good chaos indicator in the context of GR since they
cannot be undone via diffeomorphism. In the exact case however, the fractal
structure is obtained numerically; as experience dictates, this is a tool to
be use with caution with Bianchi IX.

The main technical tool of our analysis is the appeal to Ashtekar's
variables, introduced in the paper of Ashtekar of 1986 \cite{Ashte1}.
Ashtekars motivation was the quantization of GR, but he himself worked in
the application of this new formalism in Bianchi cosmologies \cite{Ashte2}.
This was not the first time however that Ashtekar's formalism was applied to
Bianchi cosmology \cite{Kodama}. The introduction of the tetrad and first
order formalisms was achieved in \cite{Samuel}. Since 1986, the ideas of
Ashtekar blossomed and ramificated in various directions as shown by the
extensive bibliography on this subject \cite{bibliography} The classical
part is treated in detail by Romano.\cite{Romano}. A short and pedagogical
introduction tailored to our needs can be found in Giulini.\cite{Giulini}.
Bianchi I and II were treated from a different point of view by Gonzalez and
Tate.\cite{Gonzalez}. Relevant treatments of the classical part can also be
found in the article of Manojlovic and Mikovic.\cite{Man-Mar} or in more
details and background in the book of Ashtekar\cite{Ashtekar}.

Our departure point is that Ashtekhar variables allow the simplification of
the hamiltonian structure of Bianchi models to the point were they can be
attacked by the methods of classical perturbation theory. These are almost
as old as Newtonian dynamics itself. One of its uses was in celestial
mechanics, where direct integration is not possible if more than two bodies
are involved. The classic perturbation method\cite{Goldstein}, based on
Hamilton-Jacobi technique shows divergences, namely,the infamous small
divisors problems\cite{L-L}. It was Poincar\'e who first realized the
complexity of the solution in phase space\cite{Poincare}. In this vein, a
great breakthrought was the Kalmagorov-Arnold-Moser (KAM) Theorem\cite{L-L}.

By now, the onset of chaos in weakly perturbed Hamiltonian systems is well
understood. In this case, the Hamiltonian splits into two parts: the
integrable part $H_0$ and the (small) perturbation $\Delta H$. Usually, one
makes a canonical transformation to action-angle variables$\left( \vec I,%
\vec \theta \right) $. Then $H=H_0\left( \vec I\right) +\Delta H\left( \vec I%
,\vec \theta \right) $. Unperturbed motion is restrained to tori on phase
space and is either periodic or quasiperiodic (integrability means that
there are $2N$ constants of motion, there being

$2N$ degrees of freedom. In the Hamiltonian case, $N$ constants of motion
are sufficient to ensure integrability since the other $N$ follow trivially
from Hamilton equations. These $N$ constants of motion are given here by $%
\vec I$, which represents the radius of these tori). Expanding $\Delta
H\left( \vec I,\vec \theta \right) $ in Fourier series

\[
\Delta H\left( \vec I,\vec \theta \right) =%
\mathop{\displaystyle \sum }
\limits_{\vec n}\Delta H\left( \vec I\right) \exp \left[ \vec n\cdot \vec 
\theta \right] 
\]

resonances will appear when $\vec N\cdot \vec \omega (I_{i0})=0$ where $\vec 
N$ are specific values for the $\vec n$, $\vec \omega $ are the unperturbed
frecuencies and $I_{i0}$ is some specific values of the actions $\vec I.$
Trying to get rid of theses primary resonances leads to secondary
resonances.By example a canonical transformation near one resonance will
lead to a Hamiltonian of the form

\[
\bar H=%
\mathop{\displaystyle \sum }
\limits_i\frac{d\omega _i(I_{i0})}{dI_i}\Delta I_i^2+\Delta H_0\cos \psi 
\]

where $\omega _i=\frac{dH(I_{i0})}{dI_i},$ $\Delta I_i=I_i-I_{i0}$ and $\psi
=\vec N\cdot \vec \theta +$ $\phi $. This is formally the Hamiltonian of the
nonlinear pendulum which its own new resonance term and its structure of
elliptic and hyperbolic fix points. That is, at a smaller scale in phase
space, a whole new resonant structure appears \cite{ZSUC}\cite{L-L}. KAM
theorem ensures that under sufficiently small perturbations, almost all tori
are preserved, but those near the resonances are destroyed. Another usual
feature is the existence of fixed point; the saddle points are of particular
interest leading to the appearance of stable and unstable manifolds and of
the so-called homoclinic (heteroclinoc if more than one saddle point are
involved) chaos\cite{L-L}.

One difficulty with the Mixmaster dynamics is that periodic or quasiperiodic
trajectories (in minisuperspace) do not exist and there is no fixed point.
The absence of periodic or cuasiperiodic solutions is due to the monotonous
increase (decrease) of the overall scale factor $\Omega \equiv \ln \left(
abc\right) $ where $a,$ $b$ and $c$ are the scale factors for the three
axes. Indeed, it has been claimed that this absence would preclude chaotic
behaviour for the Mixmaster\cite{Cushman}. But the monotonic increase
(decrease) can be easily separated from the behaviour of the other
significant variables and so one can recover quasiperiodic motions\cite{C-L}%
. Solutions corresponding to a finite number of bounces before the
trayectories make the perfect hit and go directly down one of the three
channels are forbiden by the dynamics\cite{Lin-Wald}.

All this paper can be seen as an exercise in classical mechanics in the
somewhat unusual context of General Relativity. Our goal is to investigate
Bianchi IX using Ashtekar's formulation. We will solve the integrable
Bianchi I and Bianchi II models using Hamilton-Jacobi. We will describe the
Kasner epochs in this formulation and latter apply simple canonical
perturbation theory viewing Bianchi IX as a perturbation of Bainchi II. We
will identify exactly where this approach breaks down, and analize the
reasons for this ''failure''.

\section{Bianchi models and Ashtekar's variables}

\subsection{Homogeneous Cosmologies}

Among the cosmological model of General Relativity, some are particularly
interesting for their mathematical simplicity and their physical interest:
the so-called Bianchi cosmological models\cite{Ryan-Shepley}. Their
simplicity resides in the fact that the spatial slice of these universe is
homogeneous. The number of freedoms then reduces drastically and Einstein
Equations become ordinary differential equations. The metric is 
\begin{equation}
ds^2=-N^2dt^2+q_{IJ}(t)\chi ^I\otimes \chi ^J\qquad ;\qquad I,J=1,2,3
\end{equation}

Often this metric is written using the Misner's parametrization $q_{IJ}=\exp
\left( -2\Omega (t)\right) \left( \exp 2\beta (t)\right) _{IJ}$ where $%
Tr(\beta )=0$ .This metric is (left) invariant under (spatial)
transformations generated by a certain group of symmetries that characterize
each specific Bianchi model. As usual, the Killing vectors $\xi _i$ are the
infinitesimal generators of the isometries on these spaces . The left
invariant vector fields $L_I$ ($\chi ^IL_J=\delta _J^I$ ) verify then 
\begin{equation}
\pounds _{\xi _J}L_I=[\xi _I,L_J]=0
\end{equation}

This also means that the Killing vectors are right invariant vector fields%
\cite{Wald}. The left invariant vectors are related to the structure
constants of the (Lie) group via 
\begin{equation}
\lbrack L_I,L_J]=C_{\quad IJ}^KL_K
\end{equation}

The $C_{\quad IJ}^K$ are the structure constants of the Lie group that
leaves the metric $q_{IJ}$ invariant. The spacetime spatial metric, that is,
the metric written in a coordinate basis, is given by $q_{\mu \nu
}=q_{IJ}\chi _\mu ^I\chi _\nu ^J$. General Relativity admits a Hamiltonian
formulation. Since we are interested in homogeneous cosmologies, we would
like to write a simpler Hamiltonian, one which is homogeneous from the
start. The Hamilton equations obtained from this simpler Hamiltonian will be
the correct ones (that is, the same as the one obtains using the full
Hamiltonian of General Relativity and demanding homogeneity afterward \cite
{McCallum}\cite{Ryan-Shepley}), only when the structure constants verify 
\begin{equation}
C_{\quad IJ}^K=\epsilon _{IJL}S^{LK}
\end{equation}

were $\epsilon _{IJK}$ is the antisymmetric tensor and $S^{IJ}$ is a
symmetric tensor. These models are called class A \cite{Ryan-Shepley}. The
simpler exemple is Bianchi I characterize by $C_{\quad IJ}^K=0.$ The metric
written in the Kasner form is: 
\begin{equation}
ds^2=t^{2p_1}\left( dx^1\right) ^2+t^{2p_2}\left( dx^2\right)
^2+t^{2p_3}\left( dx^3\right) ^2
\end{equation}
with $p_1+p_2+p_3=1$ and $p_1^2+p_2^2+p_3^3=1$. It is useful to use the
following parametrization for the p's 
\begin{eqnarray}
p_1 &=&\frac{-u}{u^2+u+1} \\
p_2 &=&\frac{1+u}{u^2+u+1} \\
p_3 &=&\frac{u\left( u+1\right) }{u^2+u+1}
\end{eqnarray}
where we assume the following ordering $p_1\leq p_2\leq p_3$ and $1\leq
u\leq \infty $.

Another case of interest is Bianchi IX, $C_{\quad IJ}^K=\epsilon _{\quad
IJ}^K$ . Most of the evolution of Bianchi IX can be seen as a succession of
Kasner epoch, that is the metric can be approximated to great accuracy by
the Kasner metric. This stem from the fact that the potential consist of
exponentially rising wall and are thus almost zero otherwise. Approximating
these walls as vertical, the change from one Kasner solution to another one
is given by the famous BKL u-map 
\begin{equation}
u_{n+1}=\left\{ 
\begin{array}{c}
u_n-1\qquad \quad \text{if \qquad }u_n\geq 2 \\ 
\left( u_n-1\right) ^{-1}\qquad \text{if \qquad }u_n\leq 2
\end{array}
\right.
\end{equation}
The $u_n>2$ case are called epoch and $u_n<2$ era. Chaotic behavior, if it
happen, would be confined in the era change. This could be resumed by the
following map 
\begin{equation}
u_{N+1}=u_N-[u_N]
\end{equation}
the Gauss map, relating one era with the next. Barrow proved this map to be
chaotic\cite{Barrow2}.

\subsection{Ashtekar formalism}

We will make a brief sketch of Ashtekar formalism. We will work mainly using
two basis for spacetime, an orthonormal basis $\left\{ e_a\right\} $ and a
coordinate basis $\left\{ \partial _\alpha \right\} $ . Greek indices refer
to coordinate (sometimes called spacetime in this context) bases and latin
indices to frames bases (sometimes called internal indices) . When taken
from the beginning of the alphabet ($\alpha ,$ $\beta ,...,a,b,...$ ) their
range is $\left\{ 0,1,2,3\right\} $ whereas for the middle of the alphabet ($%
\mu ,\nu ,...,i,j,...$ ) their range is $\left\{ 1,2,3\right\} $ .

The orthonormal basis and the metric are related through

\begin{eqnarray*}
\eta _{ab}e^a\otimes e^b &=&g_{\alpha \beta }dx^\alpha \otimes dx^\beta \\
\eta _{ab}e_\alpha ^ae_\beta ^b &=&g_{\alpha \beta }
\end{eqnarray*}

Thus taking determinant ont both sides gives 
\[
e^2\equiv \left( \det \left( e_\alpha ^a\right) \right) ^2=\det \left(
g_{\alpha \beta }\right) \equiv g 
\]

viewing $e_\alpha ^a$ as a square matrix of order $n$ .We will fix the
dimensionality of spacetime $n=4$ . Let's fix an orthonormal base $\left\{
^{(4)}e_{\perp },^{(4)}e_i\right\} $ so that $^{(4)}e_{\perp }$ is normal to
the hypersurface $\Sigma $. The dual co-tetrad is $\left\{ ^{(4)}e^{\perp
},^{(4)}e^i\right\} $ . We have 
\[
\frac \partial {\partial t}=Ne_{\perp }+N^ie_i 
\]

where $N$ is the lapse function and $N^ie_i$ is the shift vector field. The
relation between the spatial coordinate and frame basis is given by 
\[
\partial _\mu =e_\mu ^ie_i\qquad ;\qquad e_i=e_i^\mu \partial _\mu 
\]

The point of departure is the action 
\begin{equation}
S=\int (e)d^4x^{+}F_{\alpha \beta }^{ab}[A]e_a^\alpha e_b^\beta
\end{equation}

where $^{+}F_{\alpha \beta }^{ab}[^{+(4)}A]=2\partial _{[\alpha }\left(
^{+(4)}A_{\beta ]}^{\quad ab}\right) +2\ ^{+(4)}A_{[\alpha \quad |c|}^{\quad
a\quad }\ ^{+(4)}A_{\beta ]\quad }^{\quad cb}$ are the self-dual two-form. $%
^{+(4)}A_\alpha ^{\quad ab}$ is a self-dual connection related to the
ordinary (meaning neither self-dual or anti-self dual) connection by (see
appendix 1) 
\begin{eqnarray}
^{+(4)}A_\alpha ^{\quad ab} &\equiv &\frac 12\left( ^{(4)}A_\alpha ^{\quad
ab}-\frac 12\epsilon _{\qquad cd}^{ab}\ ^{(4)}A_\alpha ^{\quad cd}\right) \\
^{+(4)}A_\alpha ^{\quad ab} &=&i^{*}\left( ^{+(4)}A_\alpha ^{\quad ab}\right)
\nonumber \\
\ &\equiv &\frac i2\epsilon _{\qquad cd}^{ab}\ ^{+(4)}A_\alpha ^{\quad cd}
\end{eqnarray}

To recover the true degrees of freedom, as in conventional General
Relativity, one makes an ADM decomposition (\cite{ADM}, see also\cite{MTW}%
for an easier introduction; in this context, see \cite{Ashtekar}). The
result is 
\begin{eqnarray}
S &=&\int \left\{ -i\epsilon _{\quad jk}^i\partial _t\left( A_\mu ^{\quad
jk}\right) \tilde E_i^\mu -D_\mu \left( i\epsilon _{\quad jk}^i\tilde E%
_i^\mu \right) A_0^{\quad jk}\right.  \nonumber \\
&&\ \left. +\underline{N}F_{\quad \mu \nu }^{ij}\tilde E_i^\mu \tilde E%
_j^\nu +iN^\mu \epsilon _{\quad jk}^iF_{\quad \mu \nu }^{jk}\tilde E_i^\nu
\right\} d^3xdt
\end{eqnarray}

where a surface term was dropped and $D_\mu $ is the covariant derivative
constructed with $A_\mu ^{\quad ij}$ which lives exclusively in the spatial
hypersurface $\Sigma $ 
\[
D_\mu V_i=\partial _\mu V_i+A_{\mu i}^{\ \quad j}V_j 
\]
$\epsilon ^{ijk}$ is the totally antysimmetric tensor density and the
following notation was introduced 
\begin{eqnarray}
\tilde E_i^\mu &\equiv &\sqrt{q}E_i^\mu =\frac 1{\det \left( e_i^\mu \right) 
} \\
\underline{N}\ &\equiv &\frac 1{\sqrt{q}}N=\det \left( e_i^\mu \right) N
\end{eqnarray}

$E_i^\mu \equiv \perp _\nu ^\mu e_i^\nu $ is the projection of the tetrad in
the hypersurface $\Sigma .$ The action is thus written explicitly in the
form $\int \left[ p\dot q-H\right] $ whereas the Hamiltonian is a sum of
constraints, since the conjugate momentum to the lapse, shift and the
time-component of the 4-connection are absent. Our variables are $A_\mu
^{\quad jk}$ and $\Pi _{\quad jk}^\mu $, the self-dual part of $-i\epsilon
_{\quad jk}^i\tilde E_i^\mu $. The constraint's equations then read 
\begin{eqnarray}
D_\mu \left( \epsilon _{\quad jk}^i\tilde E_i^\mu \right) &=&0 \\
F_{\quad \mu \nu }^i\tilde E_i^\nu &=&0 \\
\epsilon _{\quad k}^{ij}F_{\quad \mu \nu }^k\tilde E_i^\mu \tilde E_j^\nu
&=&0
\end{eqnarray}

To write the constraints as above, we used the fact that the self-dual
Lorentz algebra is isomorphic to the Lie algebra of complexified $SO\left(
3\right) .$ The isomorphism can be carried out by taking an internal vector $%
n^a$ with $n^an_a=-1$. Then every self-dual internal 2-form $f_{ab}$ can be
characterized completely by its ''electric'' part $2f_{ab}n^a$ which is an
internal vector orthogonal to $n^a$. Hence 
\begin{eqnarray}
A_\alpha ^a &=&2iA_\alpha ^{\perp b} \\
E_a^\alpha &=&2\Pi _{\perp b}^\alpha \\
F_{\quad \mu \nu }^a &=&2iF_{\quad \text{ }\mu \nu }^{\perp a}
\end{eqnarray}

and we thus have 
\begin{equation}
F_{\quad \mu \nu }^i=2\partial _{[\mu }A_{\nu ]}^{\quad i}+\epsilon _{\quad
jk}^iA_{[\mu }^{\quad j}A_{\nu ]}^{\quad k}
\end{equation}

To connect this formalism with Bianchi cosmologies let us introduce an
orthonormal triad $E_i$ and its co-triad basis $E^i$. 
\begin{equation}
dl^2=\delta _{ij}E^iE^j
\end{equation}

Our main interest in the introduction of this basis is to use them to define
the Ashtekar's variables $\left( \tilde E_i^\mu ,A_\mu ^i\right) $%
\begin{eqnarray}
\tilde E_i^\mu &\equiv &\sqrt{\det \left( q_{\mu \nu }\right) }E_i^\mu \\
A_\mu ^i &\equiv &\Gamma _\mu ^i-iK_\mu ^i
\end{eqnarray}

The tilde denotes a tensor density. The $\Gamma _\mu ^i$ are related to the
Ricci rotation coefficients and the $K_\mu ^i$ to the extrinsic curvature.
Note the presence of $i.$By construction, Ashtekar's variables (at least in
this formulation\cite{Barbero}) are complex. We thus deal with a complex
extension of General Relativity. In the end we will need to apply some
''reality conditions '' to ensure that the (spatial) metric constructed from
the triad is real. Moreover one should also ask that the time evolution does
not make the metric complex, so that the time derivative of the metric
should also be real. Usually these variables are fields, that is they depend
on $\vec x$ as well as on time $t$ . In the homogeneous case, they will
depend only on $t$ . We can expand the triads on the invariant basis as
follow 
\begin{eqnarray}
\tilde E_i^\mu &=&E_i^IL_I^\mu |\chi | \\
A_\mu ^i &=&A_I^i\chi _\mu ^I
\end{eqnarray}

were $|\chi |\equiv \det (\chi _\mu ^I)$ . Note that in this notation $%
E_i^I=E^{Ij}\delta _{ij}$ and $E_i^I=E_{Ji}q^{IJ}$ that is capital latin
indices are raised and lowered using the invariant metric $q_{IJ}$ and the
lower case latin indices using the orthonormal metric $\delta _{ij}$ .We
have the following relation 
\begin{equation}
E_{Ii}E_J^i=q_{IJ}\det (E_i^I)
\end{equation}

where $\det (E_i^I)=\det (q_{IJ})$ .

We will be particularly interested in the diagonal case: $E_i^I\sim \delta
_i^I$ and similarly for $A_I^i$. We will denote the diagonal variables as $%
E_I$ and $A_I$ respectively. The Hamiltonian constraint $H=0$ reads 
\begin{eqnarray}
0 &=&\epsilon _i^{\quad jk}\left( -A_I^iC_{\quad JK}^I+\epsilon _{\quad
lm}^iA_J^lA_K^m\right) E_j^JE_k^K  \nonumber \\
\ &=&%
\mathop{\displaystyle \sum }
\limits_{J,K}\epsilon _i^{\quad jk}\left( -A_I^{}C_{\quad JK}^I+\epsilon
_{\quad jk}^iA_J^{}A_K^{}\right) E_J^{}E_K^{}
\end{eqnarray}

We have the following relations 
\begin{eqnarray}
K_I &\equiv &K_I^i=\ \ \frac 1{2\alpha }\frac \partial {\partial t}\left(
\ln \left( q_I\right) \right) \frac 1{E_I}\delta ^{Ii} \\
\Gamma _I &\equiv &\Gamma _I^i=\ -\ \epsilon ^{NLP}\delta _{PI}\frac 1{E_P}%
\left( \frac 12C_{\text{ }NLP}^{}-\frac 14C_{PNL}^{\quad }\right) \delta
^{Ii} \\
&&\ \ \left[ =-\ \sum\limits_{I=1}^3\epsilon ^{NLI}\frac 1{E_I}\left( \frac 1%
2C_{\text{ }NLI}^{}-\frac 14C_{INL}^{\quad }\right) \right]  \nonumber \\
A_I &\equiv &A_I^i=\frac{\omega _I}{E_I}\delta ^{Ii}=\Gamma _I-iK_I \\
E_I &=&\sqrt{\frac{\det \left( q_I\right) }{q_I}}
\end{eqnarray}

The last relation can be easily inverted to give 
\begin{equation}
q_D=\left| \frac{E_1E_2E_3}{E_D^2}\right|
\end{equation}

\section{Bianchi I}

As an introductory exercice and to explicitly relate this formalism to the
usual Kasner solution, we will solve first the Bianchi I case. It is the
simplest one, since all $C_{\quad IJ}^K=0$ . Thus, the Hamiltonian is given
by 
\begin{equation}
H=\left( \ A_1A_2E_1E_2+A_1A_3E_1E_3+A_2A_3E_2E_3\right)
\end{equation}

To find the equations of motion, in this case we could integrate quite
easily the Hamilton equations. Instead, as a preparation fot the more
complex case we will find a convenient canonical transformation that makes
the integration trivial. Consider the following generating function that
implement a change from the old (phase space) coordinates $\left( \vec A,%
\vec E\right) $ to the new ones $\left( \vec \beta ,\vec \omega \right) $: 
\begin{equation}
S(\vec A,\vec \omega )=\omega _1\ln A_1+\omega _2\ln A_2+\omega _3\ln A_3-Et
\end{equation}

Thus 
\begin{eqnarray}
E_i &=&\frac{\partial S}{\partial A^i}=\frac{\omega _i}{A^i}\text{ } \\
\beta ^i &=&\frac{\partial S}{\partial \omega _i}=\ln A^i-\frac{\partial E}{%
\partial \omega _i}t
\end{eqnarray}

and 
\begin{equation}
H=\left( \omega _1\omega _2+\omega _1\omega _3+\omega _2\omega _3\right)
\end{equation}

implying that $\omega _i$ are constants of motion and the $\beta ^i$ are
linear in the time parameter. Inverting we find 
\begin{eqnarray}
E_1 &=&\omega _1\exp \left( -\beta _1\right) \exp \left[ -(\omega _2+\omega
_3)t\right] \\
E_2 &=&\omega _2\exp \left( -\beta _2\right) \exp \left[ -(\omega _1+\omega
_3)t\right] \\
E_3 &=&\omega _3\exp \left( -\beta _3\right) \exp \left[ -(\omega _1+\omega
_2)t\right]
\end{eqnarray}

Then 
\begin{eqnarray}
q_{11} &=&\frac{\omega _2\omega _3}{\omega _1}\exp \left( \beta _1-\beta
_2-\beta _3\right) \exp \left( -2\omega _1\right) t \\
q_{22} &=&\frac{\omega _1\omega _3}{\omega _2}\exp \left( \beta _2-\beta
_1-\beta _3\right) \exp \left( -2\omega _2\right) t \\
q_{33} &=&\frac{\omega _1\omega _2}{\omega _3}\exp \left( \beta _3-\beta
_1-\beta _2\right) \exp \left( -2\omega _3\right) t
\end{eqnarray}

Up to now, our solution is complex. We have to apply the reality conditions.
This means that both the metric and its time derivative must be real. This
is easily seen to be obtained if 
\begin{equation}
\omega _i=i\Omega _i
\end{equation}

where $\Omega $ is real and 
\begin{eqnarray}
t &=&%
\mathop{\rm Re}
[t]+i%
\mathop{\rm Im}
[t] \\
&=&t_0+i\ 
\mathop{\rm Im}
[t]
\end{eqnarray}

$t_0$ a constant. We then obtain 
\begin{eqnarray}
q_{11} &=&\frac{\Omega _2\Omega _3}{\Omega _1}\exp \left[ 2\Omega _1\tilde t%
\right] \\
q_{22} &=&\frac{\Omega _1\Omega _3}{\Omega _2}\exp \left[ 2\Omega _2\tilde t%
\right] \\
q_{33} &=&\frac{\Omega _1\Omega _2}{\Omega _3}\exp \left[ 2\Omega _3\tilde t%
\right]
\end{eqnarray}

where we write $\tilde t\equiv 
\mathop{\rm Im}
[t]$ \thinspace and the prefactor can be absorbed upon a rescaling of the
axes. Also, General Relativity imposes 
\begin{equation}
H=0
\end{equation}

then to recover General Relativity, the $\Omega $ 's should obey the
following constraint: 
\begin{equation}
\Omega _1\Omega _2+\Omega _1\Omega _3+\Omega _2\Omega _3=0
\end{equation}

For convenience, to connect with the usual analysis in term of bounces and
eras later on\cite{BKL}, let us parametrize the $\Omega _i$ as follows 
\begin{eqnarray}
\Omega _1 &=&p_0+p_{+}+\sqrt{3}p_{-} \\
\Omega _2 &=&p_0+p_{+}-\sqrt{3}p_{-} \\
\Omega _3 &=&p_0-2p_{+}
\end{eqnarray}

Using the Hamiltonian constraint, we find 
\begin{equation}
p_{+}^2+p_{-}^2=p_0^2
\end{equation}

Let's introduce the angle $\theta $ : 
\begin{eqnarray}
p_{+} &=&p_0\cos \theta \\
p_{-} &=&p_0\sin \theta
\end{eqnarray}

Let us define 
\begin{eqnarray}
\sigma _1 &\equiv &\frac{\Omega _1}{\Omega _1+\Omega _2+\Omega _3}  \nonumber
\\
&=&\frac 23\left( \frac 12-\cos \left( \theta +\frac{2\pi }3\right) \right)
\\
\sigma _2 &\equiv &\frac{\Omega _2}{\Omega _1+\Omega _2+\Omega _3}  \nonumber
\\
&=&\frac 23\left( \frac 12-\cos \left( \theta -\frac{2\pi }3\right) \right)
\\
\sigma _3 &\equiv &\frac{\Omega _3}{\Omega _1+\Omega _2+\Omega _3}  \nonumber
\\
&=&\frac 23\left( \frac 12-\cos \theta \right)
\end{eqnarray}

Note that $\sigma _1+\sigma _2+\sigma _3=1$ and $\sigma _1^2+\sigma
_2^2+\sigma _3^3=1$. Redefining time 
\begin{equation}
\tilde t\equiv \frac 1{\Omega _1+\Omega _2+\Omega _3}\ln T
\end{equation}

We obtain 
\begin{eqnarray}
q_{11} &=&\frac{\Omega _2\Omega _3}{\Omega _1}T^{2\sigma _1} \\
q_{22} &=&\frac{\Omega _1\Omega _3}{\Omega _2}T^{2\sigma _2} \\
q_{33} &=&\frac{\Omega _1\Omega _2}{\Omega _3}T^{2\sigma _3}
\end{eqnarray}

Which, upon rescaling of the coordinates $x^1\rightarrow \grave x^1=\Omega
_1^{-1}\Omega _2\Omega _3\ x$ $^1$ , etc gives the well known Kasner's line
element 
\begin{equation}
ds^2=T^{2\sigma _1}\left( dx^1\right) ^2+T^{2\sigma _2}\left( dx^2\right)
^2+T^{2\sigma _3}\left( dx^3\right) ^2
\end{equation}
It is useful to divide the domain of the angle $\theta $ en six sectors 
\begin{equation}
(j-1)\frac \pi 3\leq \theta \leq j\frac \pi 3\qquad ;\qquad j\in
\{1,2,3,4,5,6\}
\end{equation}

Another parametrisation of the $\sigma _i$ is 
\begin{equation}
\tilde \sigma _1=\frac{-u}{u^2+u+1}\quad ;\quad \tilde \sigma _2=\frac{1+u}{%
u^2+u+1}\quad ;\quad \tilde \sigma _3=\frac{u(u+1)}{u^2+u+1}
\end{equation}

where $1\leq u\leq \infty $ and we write $\tilde \sigma $ to indicate the
ordering $\tilde \sigma _1\leq \tilde \sigma _2\leq \tilde \sigma _3$ . It
is straigtforward to write the $u=u(\theta )$ . The result is 
\begin{eqnarray}
j &=&1\qquad ,\qquad u=\frac{1-\cos \left( \theta +\pi /3\right) }{\cos
\theta -1/2} \\
j &=&2\qquad ,\qquad u=\frac{1+\cos \theta }{\cos \left( \theta -2\pi
/3\right) -1/2}
\end{eqnarray}

where the $j$ even (odd) formulae are generated each from the preceeding one
using $u(\theta )\rightarrow \bar u(\theta )=u\left( \theta -2\pi /3\right) $%
.

\section{Bianchi II}

This model is characterized by $C_{\quad JK}^I=\delta _3^I\epsilon _{3JK}$ .
The Hamiltonian is thus given by 
\begin{equation}
H=\left( \ A_1A_2E_1E_2+A_1A_3E_1E_3+A_2A_3E_2E_3\right) -\frac 1GA_3E_1E_2
\end{equation}

We introduce explicitly the Newton constant $G$. When this is done we have 
\cite{Ashtekar2} 
\begin{eqnarray*}
A_\mu ^i &\equiv &\frac 1G\left( \Gamma _\mu ^i-iK_\mu ^i\right) \\
F_{\quad \mu \nu }^i &=&2\partial _{[\mu }A_{\nu ]}^{\quad i}+G\ \epsilon
_{\quad jk}^iA_{[\mu }^{\quad j}A_{\nu ]}^{\quad k}
\end{eqnarray*}

As in the Bianchi I case, we begin with a canonical transformation generated
by 
\begin{equation}
F_2=\ln \left[ (GA_1)^{(G\omega _1)}(GA_2)^{(G\omega _2)}(GA_3)^{(G\omega
_3)}\right]  \label{F2}
\end{equation}

thus

\begin{equation}
H=\left( \ \omega _1\omega _2+\omega _1\omega _3+\omega _2\omega _3\right)
-\omega _1\omega _2\exp \frac{\left( \beta _3-\beta _1-\beta _2\right) }G\ 
\end{equation}

Since every one dimensional problem is integrable and since the $\beta $
appears in a very special manner in the Hamiltonian, we propose a second
canonical transformation, this time generated by 
\begin{equation}
G_2=\beta _1I_1+\beta _2I_2+\left( \beta _3-\beta _1-\beta _2\right) I_3
\label{G2}
\end{equation}

We will then go from $\left( \vec \beta ,\vec \omega \right) \rightarrow
\left( \vec \gamma ,\vec I\right) $. That will effectively convert the
hamiltonian to an equivalent unidimensional one 
\begin{equation}
H=\left( I_1I_2-I_3^2\right) -\left[ I_1I_2-\left( I_1+I_2\right)
I_3+I_3^2\right] \exp \frac{\gamma _3}G
\end{equation}

Again, with an eye on future complications, we obtain the last canonical
transformation to reduce this problem to a trivial one by applying the well
know Hamilton-Jacobi technique. The generating function that will finally
resolve our system, $S$ \cite{Landau1} 
\begin{equation}
S=W(\gamma _3,P_1,P_2,E)+\gamma _1P_1+\gamma _2P_2-Et  \label{S}
\end{equation}

is a solution of the Hamilton-Jacobi equation

\begin{equation}
H\left( P_1,P_{2,}\frac{\partial W}{\partial \gamma _3},\gamma _3\right) +%
\frac{\partial S}{\partial t}=0
\end{equation}

that is 
\begin{eqnarray}
E &=&\left( P_1P_2-\left( \frac{\partial W}{\partial \gamma _3}\right)
^2\right) \\
&&\ -\left[ P_1P_2-\left( P_1+P_2\right) \frac{\partial W}{\partial \gamma _3%
}+\left( \frac{\partial W}{\partial \gamma _3}\right) ^2\right] \exp \frac{%
\gamma _3}G
\end{eqnarray}

the other factors $P_1$and $P_2$ are constants. The solution is 
\begin{eqnarray}
W\ &=&\frac G2\left( P_1+P_2\right) \int \frac{dx}{\left( x+1\right) }-\frac 
G2\epsilon \int \frac{\sqrt{cx^2+bx+a}}{x\left( x+1\right) }dx \\
&&  \nonumber
\end{eqnarray}

where $\epsilon $ is $\pm 1$ 
\begin{eqnarray}
c &\equiv &\left( P_1-P_2\right) ^2 \\
b &\equiv &-4E \\
a &\equiv &4(P_1P_2-E)
\end{eqnarray}

and we define 
\begin{equation}
x\equiv \exp \frac{\gamma _3}G
\end{equation}

Now 
\begin{eqnarray}
\theta _3 &=&\frac{\partial W}{\partial E}-t \\
&=&\frac{-G}{\sqrt{a}}\epsilon \ \ln \left[ \frac{2\sqrt{a}\sqrt{R(x)}+bx+2a%
}{\sqrt{4ac-b^2}x}\right] -t
\end{eqnarray}

where we write $R(x)=cx^2+bx+a$ to abreviate the notation. Thus 
\begin{equation}
K\exp \left( -\sqrt{a}\epsilon \frac{\left( t+\theta _3\right) }G\right) =%
\frac{2\sqrt{a}\sqrt{R(x)}+2a+bx}x
\end{equation}

with 
\begin{eqnarray}
K\equiv \sqrt{4ac-b^2}
\end{eqnarray}

Note that $K$ posses units of $E$ (same as units of $I_i^2$ ) . Writing 
\begin{equation}
u\equiv \exp \left( -\sqrt{a}\epsilon \frac{\left( t+\theta _3\right) }G%
\right)
\end{equation}

and solving for $x$, we obtain 
\begin{equation}
x=\frac{4aKu}{\left( Ku-b\right) ^2-4ac}
\end{equation}

We can now compute $I_3$ 
\begin{eqnarray}
I_3 &=&\frac{\partial W}{\partial \gamma _3} \\
&=&\frac{-1}2\epsilon \sqrt{a}\frac{\left( Ku-4\epsilon \gamma \right)
\left( Ku+4\epsilon \lambda \right) }{\left( Ku+4\gamma \right) \left(
Ku+4\lambda \right) }
\end{eqnarray}

where we use the definitions 
\begin{eqnarray}
\gamma &=&(2P_1P_2-E)+\left( P_1+P_2\right) \sqrt{P_1P_2-E}  \label{gamma} \\
\lambda &=&\left( 2P_1P_2-E\right) -\left( P_1+P_2\right) \sqrt{P_1P_2-E}
\label{lambda}
\end{eqnarray}

The next step is to find $\theta _1$ and $\theta _2$. 
\begin{eqnarray}
\theta _1 &=&\frac{\partial S}{\partial P_1} \\
&=&\gamma _1+G\ln \frac{\left( Ku+4\gamma \right) }{\left( Ku+4\beta \right) 
}+\frac{GP_2}{\sqrt{a}}\ln u
\end{eqnarray}

similarly 
\begin{equation}
\theta _2=\gamma _2+G\ln \frac{\left( Ku+4\gamma \right) }{\left( Ku-4\alpha
\right) }+\frac{GP_1}{\sqrt{a}}\ln u
\end{equation}

were we make the choice $\epsilon =-1$ (the $\epsilon =-1$ case leads to an
equivalent solution) and we used the following definitions

\begin{eqnarray}
\alpha &=&\left( P_1-P_2\right) \sqrt{P_1P_2-E}-E  \label{alpha} \\
\beta &=&\left( P_1-P_2\right) \sqrt{P_1P_2-E}+E  \label{beta}
\end{eqnarray}

We are now able to compute the original Ashtekar's variables. Let us then
compute the $E_i$.

\begin{eqnarray}
E_1 &=&\frac 1{4G}\left( P_1+\frac 12\sqrt{a}\right) \exp \left( -\frac{%
\theta _1}G\right) \exp \left( -\frac{P_2(t+\theta _3)}G\right) \\
E_2 &=&\frac 1{4G}\left( P_2+\frac 12\sqrt{a}\right) \exp \left( -\frac{%
\theta _2}G\right) \exp \left( -\frac{P_1(t+\theta _3)}G\right) \\
E_3 &=&-\frac 1{8G\sqrt{a}}\frac{\left( K^2u^2-16\gamma ^2\right) }{Ku}\times
\nonumber \\
&&\times \exp \left( -\left( \frac{\theta _1+\theta _2}G\right) \right) \exp
\left( -\left( \frac{P_1+P_2}G\right) (t+\theta _3)\right)
\end{eqnarray}

Writing down the metric is straightforward, since $q_{11} =\frac{E_1E_2E_3}{%
E_1^2}$, and similarly the other components.

Up to now, all our work was made with the asumption that the variables were
complex. To connect with ordinary general relativity, we have to ask that $%
q_{ii}$ and $\dot q_{ii}$ should be real. But the $A^i$ should be complex
since $\Gamma ^i-iK^i$. Let us then choose 
\begin{eqnarray}
P_1 &\equiv &ip_1\qquad ,\qquad p_1\in 
\mathop{\rm Re}
{}^{+} \\
P_2 &\equiv &ip_2\qquad ,\qquad p_2\in 
\mathop{\rm Re}
\nolimits^{+} \\
E,K &\in &%
\mathop{\rm Re}
\nolimits^{+}
\end{eqnarray}

The hamiltonian constraint forces us to choose 
\begin{equation}
E=0
\end{equation}

thus

\begin{eqnarray}
\sqrt{a} &=&2\sqrt{P_1P_2} \\
\alpha &=&\sqrt{P_1P_2}\left( P_1-P_2\right) \\
\gamma &=&\sqrt{P_1P_2}(\sqrt{P_1}+\sqrt{P_2})^2
\end{eqnarray}

To understand qualitatively the evolution let us focus on the essential
features of the dynamics. Without the multiplicative constants, we have: 
\begin{eqnarray}
E_1 &=&\exp \left( -\frac{p_2t}G\right) \\
E_2 &=&\exp \left( -\frac{p_1t}G\right) \\
E_3 &=&\frac{\left( K^2u^2-16\gamma ^2\right) }{Ku}\exp \left( -\left( \frac{%
p_1+p_2}G\right) t\right)
\end{eqnarray}

where $u\equiv \exp \left( -\sqrt{a}G^{-1}t\right) $ . In the limit $%
t\rightarrow -\infty$ , we have 
\begin{eqnarray}
E_1 &=&\exp \left( -\frac{p_2t}G\right) \\
E_2 &=&\exp \left( -\frac{p_1t}G\right) \\
E_3 &=&\exp \left( -\frac{\sqrt{p_1p_2}}Gt\right) \exp \left( -\left( \frac{%
p_1+p_2}G\right) t\right)
\end{eqnarray}

In the limit $t\rightarrow \infty $ , we have 
\begin{eqnarray}
E_1 &=&\exp \left( -\frac{p_2t}G\right) \\
E_2 &=&\exp \left( -\frac{p_1t}G\right) \\
E_3 &=&\exp \left( \frac{\sqrt{p_1p_2}}Gt\right) \exp \left( -\left( \frac{%
p_1+p_2}G\right) t\right)
\end{eqnarray}

These asymptotic forms for $E_i$ have the typical aspect of a Bianchi I
solution 
\begin{eqnarray}
E_1 &=&\exp \left[ -(\Omega _2+\Omega _3)t\right] \\
E_2 &=&\exp \left[ -(\Omega _1+\Omega _3)t\right] \\
E_3 &=&\exp \left[ -(\Omega _1+\Omega _2)t\right]
\end{eqnarray}

and 
\begin{eqnarray}
E_1 &=&\exp \left[ -(\Omega _2^N+\Omega _3^N)t\right] \\
E_2 &=&\exp \left[ -(\Omega _1^N+\Omega _3^N)t\right] \\
E_3 &=&\exp \left[ -(\Omega _1^N+\Omega _2^N)t\right]
\end{eqnarray}

thus we can approximate the evolution as a transition from one Bianchi
I-like state to another (the famous ''bounce'' in the usual minisuperspace
version of homogeneous cosmology, where the evolution is seen as the motion
of a particle in a potential \cite{Ryan-Shepley}). The transition from one
Bianchi I history to another is better understood if one redefines new $%
\tilde \Omega _i$ and $\tilde \Omega _i^N$ such that 
\begin{eqnarray}
\tilde \Omega _1 &\leq &\tilde \Omega _2\leq \tilde \Omega _3 \\
\tilde \Omega _1^N &\leq &\tilde \Omega _2^N\leq \tilde \Omega _3^N
\end{eqnarray}

Then 
\begin{eqnarray}
\Omega _1^N &=&\Omega _1+2\Omega _3 \\
\Omega _2^N &=&\Omega _2+2\Omega _3 \\
\Omega _3^N &=&-\Omega _3
\end{eqnarray}

Introducing a new angle $\theta ^N$%
\begin{eqnarray}
p_{+}^N &=&p_0^N\cos \theta ^N \\
p_{-}^N &=&p_0^N\sin \theta ^N
\end{eqnarray}

we can easily find the transformation law for the parameter $\theta $%
\begin{eqnarray}
\cos \theta ^N &=&\frac{4-5\cos \theta }{5-4\cos \theta } \\
\sin \theta ^N &=&\frac{3\sin \theta }{5-4\cos \theta }
\end{eqnarray}

This is of course essentially the BKL \cite{BKL} analysis seen in a somewhat
unfamiliar context.

\section{Bianchi IX}

In this case, we will see that a straightforward application of classical
perturbation theory fails to find a canonical transformation that makes the
Hamiltonian trivial and thus the integration possible. This failure is of
course perfectly understandable in the light of recent developments
concerning Bianchi IX\cite{LMC}\cite{CGR2}\cite{C-L}.The Hamiltonian is 
\begin{eqnarray}
H &=&\left( \ A_1A_2E_1E_2+A_1A_3E_1E_3+A_2A_3E_2E_3\right) -  \nonumber \\
&&\frac 1G(A_3E_1E_2+A_1E_2E_3+A_2E_3E_1)
\end{eqnarray}

By inspection, one recognizes part of this Hamiltonian as the Bianchi II
that was shown previously to be integrable. Using the same canonical
transformation as before, namely equations (\ref{F2}) and (\ref{G2}), we
find 
\begin{eqnarray}
H &=&\left( I_1I_2-I_3^2\right) -\left[ I_1I_2-\left( I_1+I_2\right)
I_3+I_3^2\right] \exp \frac{\gamma _3}G  \nonumber \\
&&-(I_1-I_3)I_3\exp \frac{-\left( 2\gamma _1+\gamma _3\right) }G%
-(I_2-I_3)I_3\exp \frac{-\left( 2\gamma _2+\gamma _3\right) }G
\end{eqnarray}

With a further transformation

\begin{equation}
S_2=W(\gamma _3,P_1,P_2,E)+\gamma _1P_1+\gamma _2P_2
\end{equation}

using (\ref{alpha}, \ref{beta}, \ref{gamma} and \ref{lambda}) one finds

\begin{eqnarray}
I_1-I_3 &=&\left( P_1+\frac 12\sqrt{a}\right) \left( \frac{Ku+4\beta }{%
Ku+4\gamma }\right) \\
I_2-I_3 &=&\left( P_2+\frac 12\sqrt{a}\right) \left( \frac{Ku-4\alpha }{%
Ku+4\gamma }\right)
\end{eqnarray}

\begin{eqnarray}
H &=&E  \nonumber \\
&&+\frac 1{8\sqrt{a}}\frac{\left( Ku-4\gamma \right) }{Ku}\left\{ \left( P_1+%
\frac 12\sqrt{a}\right) \left( Ku-4\alpha \right) \exp \frac{-2\theta _1}G%
\exp \frac{-2P_2\theta _3}G\right. \\
&&\left. +\left( P_2+\frac 12\sqrt{a}\right) \left( Ku+4\beta \right) \exp 
\frac{-2\theta _2}G\exp \frac{-2P_1\theta _3}G\right\}  \nonumber
\end{eqnarray}

We know that General Relativity will require $H=0$. Since we want to
consider a weak perturbation, we can set $E=0$ in the perturbed Hamiltonian. 
\begin{eqnarray}
\sqrt{a} &=&2\sqrt{P_1P_2} \\
\alpha &=&\sqrt{P_1P_2}\left( P_1-P_2\right) \\
\gamma &=&\sqrt{P_1P_2}(\sqrt{P_1}+\sqrt{P_2})^2 \\
u &=&\exp \left( -2\sqrt{P_1P_2}\frac{\theta _3}G\right) \\
K &=&4\sqrt{P_1P_2}\left( P_1-P_2\right)
\end{eqnarray}

Thus 
\begin{eqnarray}
H &=&E+\frac 14\left\{ \left[ \sqrt{P_1}\left( \sqrt{P_1}+\sqrt{P_2}\right)
\left( P_1-P_2\right) u-2P_1\left( \sqrt{P_1}+\sqrt{P_2}\right) ^2\right.
\right.  \nonumber \\
&&\ \left. +\sqrt{P_1}\left( \sqrt{P_1}+\sqrt{P_2}\right) ^3u^{-1}\right]
\exp \frac{-2\theta _1}G\exp \frac{-2P_2\theta _3}G  \label{Ham} \\
&&\ +\left[ \sqrt{P_2}\left( \sqrt{P_1}+\sqrt{P_2}\right) \left(
P_1-P_2\right) u-2P_2\left( \sqrt{P_1}+\sqrt{P_2}\right) ^2\right.  \nonumber
\\
&&\ \left. \left. -\sqrt{P_2}\left( \sqrt{P_1}+\sqrt{P_2}\right)
^3u^{-1}\right] \exp \frac{-2\theta _2}G\exp \frac{-2P_1\theta _3}G\right\} 
\nonumber
\end{eqnarray}

We then have the following Hamiltonian 
\begin{equation}
H(\vec P,\vec \theta )=H_0(\vec P)+\bigtriangleup H(\vec P,\vec \theta )
\end{equation}

We are looking for a generating function $F$: 
\begin{equation}
F=\vec \theta \cdot \vec J+X(\vec \theta ,\vec J)
\end{equation}

so that 
\begin{equation}
H(\vec P,\vec \theta )\rightarrow H_0(\vec J)
\end{equation}

up to first order. Thus $F$ generates a canonical transformation from $%
\left( \vec \theta ,\vec P\right) $ to $\left( \vec \phi ,\vec J\right) $.We
want to integrate the system up to first order in the perturbation. We know
that 
\begin{eqnarray}
P_i &=&\frac{\partial F_2}{\partial \theta ^i}=J_i+\frac{\partial X}{%
\partial \theta ^i} \\
\phi ^i &=&\frac{\partial F_2}{\partial J_i}=\theta ^i+\frac{\partial X}{%
\partial J_i}
\end{eqnarray}

By Taylor 
\begin{eqnarray}
H(\vec P,\vec \theta ) &=&H(\vec J,\vec \phi )+\frac{\partial H}{\partial
\theta ^i}\left( \theta ^i-\phi ^i\right) +\frac{\partial H}{\partial P_i}%
\left( P_i-J_i\right) +O\left( \epsilon ^2\right) \\
&=&H_0(\vec J)+\bigtriangleup H(\vec J,\vec \phi )+\frac{\partial H_0}{%
\partial I_i}\frac{\partial X}{\partial \theta ^i}+O\left( \epsilon ^2\right)
\end{eqnarray}

Our case is even simpler since 
\begin{equation}
H_0=E\left( \equiv P_3\right)
\end{equation}

Thus 
\begin{equation}
H(\vec P,\vec \theta )=H_0(\vec J)+\bigtriangleup H(\vec J,\vec \phi )+\frac{%
\partial X}{\partial \theta _3}+O\left( \epsilon ^2\right)
\end{equation}

If $X$ is so that 
\begin{equation}
H(\vec P,\vec \theta )\rightarrow H_0(\vec J)+O\left( \epsilon ^2\right)
\end{equation}

then 
\begin{equation}
\bigtriangleup H(\vec J,\vec \phi )+\frac{\partial X}{\partial \theta _3}%
=O\left( \epsilon ^2\right)
\end{equation}

that is 
\begin{equation}
\frac{\partial X}{\partial \theta _3}=-\bigtriangleup H(\vec J,\vec \phi
)+O\left( \epsilon ^2\right)
\end{equation}

which is the same as asking that 
\begin{equation}
\frac{\partial X}{\partial \theta _3}=-\bigtriangleup H(\vec J,\vec \theta
)+O\left( \epsilon ^2\right)
\end{equation}

or 
\begin{equation}
X=-\int \bigtriangleup H(\vec J,\vec \theta )d\theta _3+O\left( \epsilon
^2\right)
\end{equation}

Thus 
\begin{eqnarray}
X &=&\frac G8\left\{ \left[ \frac{\sqrt{P_1}\left( P_1-P_2\right) }{\sqrt{P_2%
}}\exp \frac{-2\sqrt{P_2}\left( \sqrt{P_1}+\sqrt{P_2}\right) \theta _3}G%
\right. \right.  \nonumber \\
&&\left. \frac{-2P_1\left( \sqrt{P_1}+\sqrt{P_2}\right) ^2}{P_2}\exp \frac{%
-2P_2\theta _3}G\right.  \nonumber \\
&&\left. -\frac{\sqrt{P_1}\left( \sqrt{P_1}+\sqrt{P_2}\right) ^3}{\sqrt{P_2}%
\left( \sqrt{P_1}-\sqrt{P_2}\right) }\exp \frac{-2\sqrt{P_2}\left( \sqrt{P_2}%
-\sqrt{P_1}\right) \theta _3}G\right] \exp \frac{-2\theta _1}G \\
&&+\left[ \frac{\sqrt{P_2}\left( P_1-P_2\right) }{\sqrt{P_1}}\exp \frac{-2%
\sqrt{P_1}\left( \sqrt{P_1}+\sqrt{P_2}\right) \theta _3}G-\frac{2P_2\left( 
\sqrt{P_1}+\sqrt{P_2}\right) ^2}{P_1}\exp \frac{-2P_1\theta _3}G\right. 
\nonumber \\
&&\left. \left. -\frac{\sqrt{P_2}\left( \sqrt{P_1}+\sqrt{P_2}\right) ^3}{%
\sqrt{P_1}\left( \sqrt{P_1}-\sqrt{P_2}\right) }\exp \frac{-2\sqrt{P_1}\left( 
\sqrt{P_1}-\sqrt{P_2}\right) \theta _3}G\right] \exp \frac{-2\theta _2}G%
\right\}  \nonumber
\end{eqnarray}

is the generating function that would permit the integration of the whole
system albeit to first order. But the generating function $X$ diverges and
is thus useless near the following planes 
\begin{eqnarray}
P_1 &=&0 \\
P_2 &=&0 \\
P_1 &=&P_2
\end{eqnarray}

The physical significance of these values is explained in the appendix 2.

Let us analyse the $P_2=0$ resonance. It is clear that the problem with $X$
arises because the exponents lead to inverse powers of $P_2$ in the
generating function; positive powers of $P_2$ in the pre exponential factors
only improve integrability; thus, to analyze whether the system may be
integrated or not, we may drop all powers of $P_2$ from the prefactors and
write $H$ as

\begin{eqnarray}
H &=&E+\frac 14P_1^2\exp \frac{-2\theta _1}G\exp \frac{-2P_2\theta _3}G%
\left\{ -2+\exp \frac{-2\sqrt{P_2}\sqrt{P_1}\theta _3}G+\exp \frac{2\sqrt{P_2%
}\sqrt{P_1}\theta _3}G\right\}  \nonumber \\
\ &=&E+\frac 12P_1^2\exp \frac{-2\theta _1}G\exp \frac{-2P_2\theta _3}G%
\left\{ \cosh \frac{2\sqrt{P_2}\sqrt{P_1}\theta _3}G-1\right\}
\end{eqnarray}

The typical (perturbative) route to chaos is seen by seeking second order
resonances. Let's then integrate the more resonant term using the following
generating function 
\begin{equation}
S_2=G\alpha \exp \left( \frac{\theta _1}G\right) +\varepsilon \theta
_2+K\theta _3-\frac 14\frac{\alpha ^2G}\varepsilon \left( \exp \left[ -2%
\frac{\varepsilon \theta _3}G\right] -1\right)  \label{action}
\end{equation}

which allows us to make the change from $\left( \theta _1,\theta _2,\theta
_3,P_1,P_2,E\right) \rightarrow \left( \psi _1,\psi _2,\psi _3,\alpha
,\varepsilon ,K\right) $ . The new Hamiltonian reads 
\begin{equation}
H=K+\frac 12\alpha _1^2\exp \frac{-2\varepsilon \psi _3}G\cosh \left\{ \frac{%
2\sqrt{\varepsilon }\psi _3}G\sqrt{\frac \alpha G\psi _1+\frac{\alpha ^2}{%
2\varepsilon }\left( \exp \frac{-2\varepsilon \psi _3}G-1\right) }\right\}
\end{equation}

where 
\begin{equation}
K=E+\frac 12P_1^2\exp \frac{-2\theta _1}G\exp \frac{-2P_2\theta _3}G
\end{equation}

While we seem to have isolated the resonance at $\varepsilon =0$, the new
Hamiltonian has other secundary resonances. To bring them forth, let us
attempt to integrate the new Hamiltonian to first orden using 
\begin{eqnarray}
X &=&-\int \triangle Kd\psi _3 \\
&=&-\frac{G\alpha ^2}{4\varepsilon }\int \exp \left( -z\right) \cosh \left\{
z\sqrt{\hat A+\hat B\exp \left( -z\right) }\right\} dz
\end{eqnarray}

where 
\begin{eqnarray}
\hat A &=&\frac \alpha {\varepsilon G}\psi _1-\frac{\alpha ^2}{2\varepsilon
^2}=\frac \alpha \varepsilon \left( \frac{\psi _1}G-\frac \alpha {%
2\varepsilon }\right) \\
\hat B &=&\frac{\alpha ^2}{2\varepsilon ^2}
\end{eqnarray}

Consider 
\begin{eqnarray}
I_{+} &=&\int \exp \left( -z\right) \exp \left( zA\sqrt{1+B\exp \left(
-z\right) }\right) dz  \nonumber \\
&=&\int \exp \left[ \left( -1+A\sqrt{1+B\exp \left( -z\right) }\right)
z\right] \\
I_{-} &=&\int \exp \left( -z\right) \exp \left( -zA\sqrt{+B\exp \left(
-z\right) }\right) dz  \nonumber \\
&=&\int \exp \left[ -\left( 1+A\sqrt{1+B\exp \left( -z\right) }\right)
z\right]
\end{eqnarray}

where$\ $%
\begin{eqnarray}
A &=&\sqrt{\hat A}=\sqrt{\frac \alpha \varepsilon \left( \frac{\psi _1}G-%
\frac \alpha {2\varepsilon }\right) }  \label{A} \\
B &=&\frac{\hat B}{\hat A}=\frac{\alpha ^{}}{2\varepsilon \left( \frac{\psi
_1}G-\frac \alpha {2\varepsilon }\right) }
\end{eqnarray}

Thus 
\begin{equation}
X=-\frac{G\alpha ^2}{8\varepsilon }\left( I_{+}+I_{-}\right)
\end{equation}

Let's analyse the divergence of $I_{\pm }$ . Let's tackle the case of $I_{+}$
first. 
\begin{equation}
I_{+}(A)=\sum\limits_{k=0}^\infty I_kA^k
\end{equation}

where 
\begin{eqnarray}
I_k &=&\frac 1{2\pi i}\oint \frac 1{A^{k+1}}\int dz\exp \left[ \left( -1+A%
\sqrt{1+B\exp \left( -z\right) }\right) z\right]  \nonumber \\
&=&\frac 1{k!}\int_0^\infty x^k\exp \left( -x\right) \left( 1+B\exp \left(
-x\right) \right) ^{k/2}dx \\
&\simeq &\left( 1+B\exp \left( -k\right) \right) ^{k/2}
\end{eqnarray}

The series diverges when $A=1$ . We argue that $A=1$ is a (simple) pole for
this series, at least for some range of the initial parameters. To show
this, we will construct an analytic continuation for the series (see
appendix 3). Then 
\begin{equation}
I_{+}=\sum_{k=0}^\infty A^k+\sum_{k=0}^\infty F_k(B)\left( \frac A2\right) ^k
\end{equation}

where 
\begin{equation}
F_k(B)\leq \frac 14B\frac{\left( \sqrt{1+B}\right) ^k-1}{\sqrt{1+B}-1}
\end{equation}

We argue that $A=1$ is a simple pole since the series 
\begin{equation}
\sum_{k=0}^\infty F_k(B)\left( \frac A2\right) ^k
\end{equation}
are analytic in $A=1$. Using equation \ref{A} we encounter ''resonance'' for
the values 
\begin{equation}
\frac \alpha \varepsilon \left( \frac{\psi _1}G-\frac \alpha {2\varepsilon }%
\right) =1
\end{equation}

That is 
\begin{equation}
\epsilon =\frac 12P_1\pm \frac 12\sqrt{P_1^2-2\alpha }
\end{equation}

where we used the generating function \ref{action} to obtain the relation
between $\alpha ,\psi _1$ and $P_1$%
\begin{equation}
\frac \alpha G\psi _1=P_1
\end{equation}

\section{Conclusion}

Using Ashtekar formalism, we treat the Bianchi IX Hamiltonian using the
tools of classical perturbation theory. Namely, we treat Bianchi IX as a
perturbation of the integrable Bianchi II. The failure to integrate Bianchi
IX perturbatively did not come as a surprise in view of the numerous
analytical and numerical evidences of its chaotic character. Our main
purpose was to show how the use of Ashtekar variable simplify the
perturbative analysis since the Hamiltonian (often refered to as the scalar
constraint in this context) is notably simpler then in the usual one. As a
by product of our analysis we filled some gaps in the literature, such us
recovering the BKL map in this context. The reality conditions, which are
often difficult to handle are easily deal with here using the known Bianchi
II solution and the relation between the two formalisms.

From the point of view of the larger framework of chaos and general
relativity, the important point is to analyze the reasons for the
''failure'' of our naive approach to Bianchi IX. Our strategy was simply to
apply to Bianchi IX, once rendered manageable by the translation into
Ashtekhars variables, the most direct known ways to handle a Hamiltonian
system. We treaded at first on known grounds, and not surprisingly found a
quick success (providing a more complete solution to the Bianchi II case
than previously reported, a tribute to the power of Hamilton - Jacobi
methods). Thus encouraged, we faced the Bianchi IX problem.

The expectation in facing a complex Hamiltonian dynamical system, but that
can be divided into an integrable part and a "perturbation", is that either
no resonances will appear, and then the system will be integrable by virtue
of KAM theory, or else there will be resonances. If these are isolated,
however, the dynamics is not yet chaotic, but rather the resonances appear
as boundaries of regions of integrability. Chaos arises when resonances
appear in layers, an infinite set of "secondary" resonances accumulating
towards the primary ones.

Because of the complex nature of Ashtekhar variables, it could not be
expected that this analysis would translate in any straightforwrad way to
our problem; rather, our aim has been to discern whether a similar structure
to the familiar weak chaos appeared in our problem. Thus, instead of primary
resonances, we found that naive perturbation theory (seeing Bianchi IX as a
perturbation of Bianchi II) breaks down in channel runs, and also near the
singular solutions at the end of the channels. The issue then became whether
these were isolated singularities, or rather there were other breakdown
points arbitrarily close to them.

In order to find an answer, we zeroed on the $P_2=0$ singular metric, by
isolating the terms responsible for the breakdown of naive perturbation
theory, and proceeded to the exact integration of this most singular part of
the Hamiltonian. If (to our surprise) no further obstructions to
integrability had appeared, this would have been an indicator of a non
chaotic nature of the whole Bianchi IX system; alas, our expectations held
on, and we found that, arbitrarily close to the original singular solution,
new singularities of perturbation theory appeared, as given by Eqs. (186)
and (187).

To analyze these conditions, we must remember that, from the point of view
of the dynamics generated by the Hamiltonian $K$, both $\alpha$ and $\psi_1$
are constants of motion. Thus what we found is, for any given value of the
original momentum $P_1$, an hyperbola of trajectories where perturbation
theory breaks down, approaching asymptotically the known singularity at $%
\varepsilon =0$ when $\alpha\to 0$. The $P_2=\varepsilon =0$ singular point
cannot be isolated, therefore, and we must expect complex behavior in a
neigborhood of the corners of Misner´s triangular potential well.

We see that the failure of our naive approach in fact is leading us directly
to the sources of complexity in the dynamics. This bolsters the conclusion
that Ashtekar variables are a useful tool in studying chaotic behavior in
General Relativity.

We are indebted to Luca Bombelli, Viqar Husain, Ted Jacobson, Jorge Pullin
and Michael Ryan for enlightening discussions.This work was partially
supported by Universidad de Buenos Aires, CONICET, Fundaci\'on Antorchas,
and by the Comission of the European Communities under contract Nr.
C11*-CJ94-0004.

\section{Appendix 1}

We will resume here some basic notions of differential geometry. Our
notation follows \cite{MTW}. First, given two n-forms $\lambda $ and $\sigma 
$ at some point in $M$ : 
\begin{eqnarray}
\lambda &=&\frac 1{n!}\lambda _{a_1...a_n}e^{a_1}\wedge ...\wedge e^{a_n} \\
\sigma &=&\frac 1{n!}\sigma _{a_1...a_n}e^{a_1}\wedge ...\wedge e^{a_n}
\end{eqnarray}

we defined their inner product, induced by $g$ , via 
\begin{equation}
<\lambda ,\sigma >\ \equiv \frac 1{n!}\lambda _{a_1...a_n}\
g^{a_1b_1}...g^{a_nb_n}\sigma _{b_1...b_n}=\lambda _{a_1...a_n}\sigma
^{a_1...a_n}
\end{equation}

The following proposition follows readily 
\begin{equation}
\lambda \wedge ^{\star }\sigma =<\lambda ,\sigma >\epsilon
\end{equation}

where $\epsilon $ is the volume form on $M$ induced by the metric 
\begin{eqnarray}
\epsilon &=&\sqrt{g}dx^0\wedge ...\wedge dx^{(n-1)} \\
&=&\frac 1{n!}\epsilon _{a_1...a_n}\ e^{a_1}\wedge ...\wedge e^{a_n}
\end{eqnarray}

where $\epsilon _{a_1...a_n}\equiv n!\delta _{[a_1}^0...\delta
_{a_n]}^{(n-1)}$ .

Consider the following string of equalities 
\begin{eqnarray}
S &=&\int F_{ab}\wedge ^{*}\left( e^a\wedge e^b\right)  \nonumber \\
&=&\frac 12\int R_{abcd}\left( e^c\wedge e^d\right) \wedge ^{*}\left(
e^a\wedge e^b\right)  \nonumber \\
&=&\frac 12\int R_{abcd}<e^c\wedge e^d,e^a\wedge e^b>\epsilon  \nonumber \\
&=&\frac 12\int R_{abcd}\left( \eta ^{ac}\eta ^{bd}-\eta ^{ad}\eta
^{bc}\right) \epsilon  \nonumber \\
&=&\int R\epsilon
\end{eqnarray}

This show how to write the Einstein-Hilbert action in terms of the curvature
2-form $F_{ab}$ and the co-tetrads $e^a$ . The curvature $F_{ab}$ is defined
as 
\begin{equation}
F_{ab}\equiv \frac 12R_{abcd}\ e^c\wedge e^d
\end{equation}

and satisfy the second Cartan 's structure equation 
\begin{equation}
d\omega _{\quad b}^a+\omega _{\quad c}^a\wedge \omega _{\quad b}^c=F_{ab}
\end{equation}

The $\omega _{\quad b}^a$ are the connection 1-forms, defined as 
\begin{equation}
\omega _{\quad b}^a\equiv \Gamma _{bc}^ae^c
\end{equation}

The component of the $\omega _{\quad b}^a$ in coordinate base are called the
Ricci rotation coefficients\cite{Wald} 
\begin{equation}
\Gamma _{bc}^ae_\alpha ^cdx^\alpha =\omega _{\alpha \quad b}^{\quad
a}dx^\alpha
\end{equation}

That is 
\begin{equation}
\omega _{\alpha \quad b}^{\quad a}=e_\alpha ^c\Gamma _{bc}^a
\end{equation}

It is straightforward to prove that $\omega _{ab}=-\omega _{ba}$ \cite{Wald}%
. The following definition is useful 
\begin{eqnarray}
\nabla _\alpha V &=&\nabla _\alpha \left( V_ae^a\right)  \nonumber \\
\ &=&\partial _\alpha \left( V_a\right) e^a+V_ae_\alpha ^b\nabla _be^a 
\nonumber \\
\ &=&\left( \partial _\alpha \left( V_a\right) -V_c\omega _{\alpha \quad
a}^{\quad c}\right) e^a  \nonumber \\
\ &=&\left( \partial _\alpha \left( V_a\right) +V_c\omega _{\alpha a}^{\quad
\ c}\right) e^a\equiv D_\alpha \left( V_a\right) e^a
\end{eqnarray}

Finally the action can be rewritten as follows 
\begin{eqnarray}
S &=&\int F_{ab}\wedge ^{*}\left( e^a\wedge e^b\right)  \nonumber \\
\ &=&\frac 12\int R_{abcd}\left( e^c\wedge e^d\right) \wedge ^{*}\left(
e^a\wedge e^b\right)  \nonumber \\
\ &=&\frac 12\int R_{ab\alpha \beta }e_c^\alpha e_d^\beta \left( e^c\wedge
e^d\right) \wedge ^{*}\left( e^a\wedge e^b\right)  \nonumber \\
\ &=&\frac 12\int R_{ab\alpha \beta }e_c^\alpha e_d^\beta <e^c\wedge
e^d,e^a\wedge e^b>\epsilon  \nonumber \\
\ &=&\frac 12\int R_{ab\alpha \beta }e_c^\alpha e_d^\beta \left( \eta
^{ac}\eta ^{bd}-\eta ^{ad}\eta ^{bc}\right) \epsilon  \nonumber \\
\ &=&\int R_{\quad \alpha \beta }^{cd}e_c^\alpha e_d^\beta (e)d^4x
\end{eqnarray}

where $d^4x$ is a short notation for $dx^0\wedge ...\wedge dx^{\left(
n-1\right) }$ .

\section{Appendix 2}

The purpose of this appendix is to give the physical meaning of the
resonances 
\begin{eqnarray}
P_1 &=&0 \\
P_2 &=&0 \\
P_1 &=&P_2
\end{eqnarray}
First let us write the metric: 
\begin{equation}
q_{11}=\frac{-1}{16P_1G}\frac{\left( Ku\right) ^2-16\gamma ^2}{Ku}\exp
\left( -2\frac{\theta _2}G\right) \exp \left( -2P_1\frac{\theta _3}G\right)
\end{equation}

\begin{equation}
q_{22}=\frac{-1}{16P_2G}\frac{\left( Ku\right) ^2-16\gamma ^2}{Ku}\exp
\left( -2\frac{\theta _1}G\right) \exp \left( -2P_2\frac{\theta _3}G\right)
\end{equation}

\begin{equation}
q_{33}=\frac{-1}{2G}P_1P_2(\sqrt{P_1}+\sqrt{P_2})^2\frac{Ku}{\left(
Ku\right) ^2-16\gamma ^2}
\end{equation}

where 
\begin{eqnarray}
u &\equiv &\exp \left( -\sqrt{P_1P_2}\frac{\theta _3}G\right) \\
\gamma &=&\sqrt{P_1P_2}(\sqrt{P_1}+\sqrt{P_2})^2
\end{eqnarray}

We can rewrite this in a more convenient manner. To recover real general
relativity, we choose $P_1$ and $P_2$ imaginary and write separately the
real and imaginary part of $\theta _3.$ We finally find 
\begin{eqnarray}
q_{11} &=&\frac{-i(\sqrt{P_1}+\sqrt{P_2})^2}{2G}\sqrt{\frac{P_2}{P_1}}\cosh
\left( \sqrt{P_1P_2}\frac{\theta _3}G\right) \exp \left( -2\frac{\theta _2}G%
\right) \exp \left( -2P_1\frac{\theta _3}G\right) \\
q_{22} &=&\frac{-i(\sqrt{P_1}+\sqrt{P_2})^2}{2G}\sqrt{\frac{P_1}{P_2}}\cosh
\left( \sqrt{P_1P_2}\frac{\theta _3}G\right) \exp \left( -2\frac{\theta _1}G%
\right) \exp \left( -2P_2\frac{\theta _3}G\right) \\
q_{33} &=&\frac 1G\frac{i\sqrt{P_1P_2}}{\,\cosh \left( \sqrt{P_1P_2}%
G^{-1}\theta _3\right) }
\end{eqnarray}

where we write $\theta _3$ instead of $\mathop{\rm Im}[\theta _3]$ to
simplify the notation. If $P_1=P_2$, $q_{11}=q_{22}$.

Recall that, in the standard Misner notation, 
\begin{eqnarray}
q_{11} &=&\exp 2\left( -\Omega +\beta _1+\sqrt{3}\beta _2\right) \\
q_{22} &=&\exp 2\left( -\Omega +\beta _1-\sqrt{3}\beta _2\right)
\end{eqnarray}

Thus 
\begin{equation}
q_{11}=q_{22}\Rightarrow \beta _2=0
\end{equation}

Viewing the evolution of the universe as a scattering process in a
triangular potential well, this gives us a particle-universe running right
into the right hand channel.

Let's see now the case $P_2\rightarrow 0$ . In this case 
\begin{equation}
q_{11}\rightarrow \infty
\end{equation}

while the other two components of the metric go to zero. This correspond to
a particle going into one of the left hand channel. Similar behavior occurs
with $P_1\rightarrow 0$.

\section{Appendix 3}

Let's go back to our exact integral for $I_k$ : 
\begin{equation}
I_k=\frac 1{k!}\int_0^\infty x^k\exp \left( -x\right) \left( 1+B\exp \left(
-x\right) \right) ^{k/2}dx
\end{equation}

We can write 
\begin{equation}
\left( 1+B\exp \left( -x\right) \right) ^{k/2}=1+\left[ \sqrt{1+B\exp \left(
-x\right) }-1\right] \sum_{i=0}^{k-1}\left( \sqrt{1+B\exp \left( -x\right) }%
\right) ^i
\end{equation}

If $|B|<1$ , then 
\begin{equation}
\sqrt{1+B\exp \left( -x\right) }\leq 1+\frac 12B\exp \left( -x\right)
\end{equation}

and 
\begin{eqnarray}
\sum_{i=0}^{k-1}\left( \sqrt{1+B\exp \left( -x\right) }\right) ^i &\leq
&\sum_{i=0}^{k-1}\left( \sqrt{1+B}\right) ^i \\
&=&\frac{\left( \sqrt{1+B}\right) ^k-1}{\sqrt{1+B}-1}
\end{eqnarray}

Thus 
\begin{eqnarray}
I_k &=&\frac 1{k!}\left\{ \int_0^\infty x^k\exp \left( -x\right)
dx+\int_0^\infty x^k\exp \left( -x\right) \left[ \sqrt{1+B\exp \left(
-x\right) }-1\right] \sum_{i=0}^{k-1}\left( \sqrt{1+B\exp \left( -x\right) }%
\right) ^idx\right\}  \nonumber \\
&=&1+\frac 1{k!}\int_0^\infty x^k\exp \left( -x\right) \left[ \sqrt{1+B\exp
\left( -x\right) }-1\right] \sum_{i=0}^{k-1}\left( \sqrt{1+B\exp \left(
-x\right) }\right) ^idx
\end{eqnarray}

now 
\begin{eqnarray*}
\int_0^\infty x^k\exp \left( -x\right) \left[ \sqrt{1+B\exp \left( -x\right) 
}-1\right] \sum_{i=0}^{k-1}\left( \sqrt{1+B\exp \left( -x\right) }\right)
^idx &\leq &\frac 14B\frac{\left( \sqrt{1+B}\right) ^k-1}{\sqrt{1+B}-1}\frac 
1{2^k}\int_0^\infty y^k\exp \left( -y\right) dy \\
&=&\frac 14B\frac{\left( \sqrt{1+B}\right) ^k-1}{\sqrt{1+B}-1}\frac 1{2^k}k!
\end{eqnarray*}

Therefore 
\begin{eqnarray}
I_k &=&1+F_k\left( B\right)
\end{eqnarray}

where 
\begin{equation}
F_k(B)\leq \frac 14B\frac{\left( \sqrt{1+B}\right) ^k-1}{\sqrt{1+B}-1}\frac 1%
{2^k}
\end{equation}

\end{document}